\documentclass[12pt,superscriptaddress,showpacs,showkeys,twocolumn,prb,aps,reprint,a4paper,floatfix]{revtex4-1}

\usepackage{graphicx}

\usepackage{amsmath, amsthm, amssymb, amsfonts}
\usepackage{float}

\usepackage{natbib}
\usepackage{physics}
\usepackage[para,online,flushleft]{threeparttable} 
\usepackage{array}
\usepackage{ragged2e} 

\makeatletter
\newcommand*{\citenst}[2][]{%
  \begingroup
  \let\NAT@mbox=\mbox
  \let\@cite\NAT@citenum
  \let\NAT@space\NAT@spacechar
  \let\NAT@super@kern\relax
  \renewcommand\NAT@open{[}%
  \renewcommand\NAT@close{]}%
  \citet[#1]{#2}%
  \endgroup
}
\makeatother

\makeatletter
\newcommand*{\citenumns}[2][]{%
  \begingroup
  \let\NAT@mbox=\mbox
  \let\@cite\NAT@citenum
  \let\NAT@space\NAT@spacechar
  \let\NAT@super@kern\relax
  \renewcommand\NAT@open{[}
  \renewcommand\NAT@close{]}%
  \cite[#1]{#2}
  \endgroup
}
\makeatother
\usepackage[caption=false]{subfig}

\begin{document}
\title{Nonlinear mechanisms in Al and Ti superconducting travelling-wave parametric amplifiers}
\author{Songyuan Zhao}
\email{sz311@cam.ac.uk}
\author{S. Withington}
\author{C. N. Thomas}
\date{February 23, 2022}

\affiliation{Cavendish Laboratory, JJ Thomson Avenue, Cambridge CB3 OHE, United Kingdom.}

\begin{abstract}
\noindent  The underlying nonlinear mechanisms behind the operation of travelling-wave parametric amplifiers (TWPAs) are important in determining their performance in terms of added noise, maximum gain, and bandwidth. We describe a method of characterising the underlying nonlinearity of a superconducting material in terms of its dissipative-reactive ratio and the response time of the underlying microscopic processes. We describe and calculate the different behaviour arising from the \textit{equilibrium supercurrent nonlinearity}, which has low dissipation and fast response time, and the \textit{non-equilibrium heating nonlinearity}, which has high dissipation and slow response time. We have fabricated TWPAs based on Al and Ti, and characterised their nonlinearities using our analysis. For both Al and Ti, the measured dissipative-reactive ratios and response times are quantitatively similar to predictions for the non-equilibrium heating nonlinearity. We were able to obtain more than $20\,\mathrm{dB}$ of peak power gain, although only over a narrow bandwidth of a few kilohertz. Our method of characterising the underlying nonlinearities could also be useful in the understanding and design of other superconducting nonlinear devices such as parametric up-converters, kinetic inductance Fourier transform spectrometers, and resonator parametric amplifiers.
\end{abstract}


\maketitle

\section{Introduction}
Superconducting travelling-wave parametric amplifiers (TWPAs) are low-noise cryogenic amplifiers based on thin-film superconducting transmission lines \citenumns{Eom_2012}. These devices have received considerable interest in the last decade due to their potential to achieve high gain, broad-band amplification whilst having noise performance close to the quantum limit \citenumns{Malnou_2020}. Currently, there are two main types of superconducting TWPAs: Josephson travelling-wave parametric amplifiers (J-TWPAs) and kinetic inductance travelling-wave parametric amplifiers (KI-TWPAs). J-TWPAs rely on the nonlinear inductance of tunnel junctions to achieve amplification. These devices benefit from the wealth of literature on designing resonant Josephson parametric amplifiers (JPAs) and Josephson junctions. However, the saturation powers of J-TWPAs are limited by the same junction critical current as their JPAs counterpart, and they may, as a result, be unsuitable for certain applications. KI-TWPAs, on the other hand, have high saturation powers, which are determined by the critical currents of the materials. They are suitable for applications such as reading out large detector or qubit arrays. In this study, we focus on the underlying nonlinearities behind the operation of KI-TWPAs.

KI-TWPAs achieve amplification through wave-mixing processes induced by the nonlinear superconducting kinetic inductance. In general, the kinetic inductance per unit length $L_{\mathrm{k}}$ of a superconducting transmission line can be expanded as an infinite series of even powers of local line current ${I}$ as follows:
\begin{align}
  L_{\mathrm{k}} = L_{\mathrm{k},0}\left(1+\frac{{I}^2}{I_*^2}+\frac{{I}^4}{I_{*,4}^4}+\cdots\right)\,, \label{eq:nonlinear_scale}
\end{align}
where $I_{*}$ and $I_{*,4}$ are the scales of the quadratic and quartic orders of nonlinearity respectively \citenumns{Eom_2012, Jonas_review, Kher_2017, songyuan2019_nonlinear}, and $L_{\mathrm{k},0}$ is the kinetic inductance per unit length in the absence of inductive nonlinearity. The odd powers of dependence on current must vanish since, for an infinitesimal segment of superconducting transmission line, the kinetic inductance has the same value whether a $+I$ current is applied or a $-I$ current is applied. This symmetry can be broken through the application of a DC bias current, i.e. $I\rightarrow I_{\mathrm{DC}}+I_{\mathrm{RF}}$, where $I_{\mathrm{DC}}$ is the DC current and $I_{\mathrm{RF}}$ is the RF current. In the presence of a DC-bias, the kinetic inductance therefore also depends on the odd powers of the RF current. As a result of this reactive kinetic inductance nonlinearity, when a strong pump current propagates with a weak signal current along a TWPA, energy is transferred from the pump to the signal, thereby achieving parametric amplification \citenumns{Eom_2012, Songyuan_2021_Thesis}. The form of inductance given in equation~(\ref{eq:nonlinear_scale}) can arise through different underlying physical contributions, such as the \textit{equilibrium supercurrent nonlinearity} and the \textit{non-equilibrium heating nonlinearity}. The meanings of these terms will be discussed in the theory section of this work.


It is important to understand the nonlinear mechanisms behind the amplification process for any specific combination of material and geometry. In particular, the dissipative aspect and the response time of a mechanism can have profound impact on the amplification performance of a device. Firstly, any increase in dissipation with applied power will decrease the power gain. Secondly, an increase in dissipated power may also cause heating of the device, thereby further degrading its performance. Thirdly, a dissipative nonlinearity is likely to alter the noise behaviour of the device, e.g. altering the thermal noise output by changing the level of dissipation, or by resulting in additional modulation of the noise. Understanding the magnitude of the dissipative nonlinearity is therefore important in determining if a particular TWPA can, in theory, achieve amplification and approach the quantum limit \citenumns{Zhao_2021}. Lastly, as we shall show in this study, the response time of a nonlinear mechanism imposes a strong constraint on the bandwidth of a TWPA. In the case of non-equilibrium heating nonlinearity, the bandwidth is restricted to the range of a few $\mathrm{kHz}$ to a few $\mathrm{MHz}$, depending on the material. This severely undermines the advantage of the broad-band travelling-wave geometry. A proper understanding of the underlying nonlinear mechanisms will inform the choice of material and the design of a TWPA so as to avoid these limitations.

In this study, we present a theory and measurements to investigate the nonlinear mechanisms of Al and Ti TWPAs. We characterise these mechanisms in terms of dissipative-reactive ratios and response times. Measurements of both parameters can be performed using the same transmission line structure as the travelling-wave amplifier itself, as opposed to requiring a separate geometry such as a resonator. We calculate the expected values of dissipative-reactive ratios of both the equilibrium supercurrent nonlinearity as well as the non-equilibrium heating nonlinearity. We also present a response time model of the nonlinear inductance, and show that the model predicts characteristic features in the measured gain profiles, such as bandwidth limitations and post-gain dips. The effects of reactive, dissipative, and rate-limited nonlinearity are also discussed extensively, in the context of resonator parametric amplifiers, in an upcoming paper \citenumns{Chris_2022_resonator_amplifier}. In the experimental section, we present measurements of Al and Ti TWPAs that we have fabricated. Our measurements show significant dissipative nonlinearities in line with the order of magnitude predicted by the non-equilibrium heating nonlinearity. Further, our measurements of gain profiles confirm the characteristic features of our response time model of nonlinear inductance. Together, these measurements demonstrate that the nonlinear responses of Al and Ti seem to contain significant contributions from non-equilibrium heating nonlinearity. In line with this, we were able to obtain more than $20\,\mathrm{dB}$ of peak power gain, although only over a narrow bandwidth of a few kilohertz. 



\section{Nonlinear mechanisms}
\subsection{Overview of mechanisms}
In this work, we focus on two important effects of currents that have been widely observed in superconducting films \citenumns{Jonas_review, Tinkham_1994, Parmenter_1962, Goldie_2012, de_Visser_2012, Visser_2014}: 1) modification of the quasiparticle density of states and therefore the equilibrium charge carrier distribution, and 2) non-equilibrium power absorption and sub-gap pair-breaking. We shall refer to the first effect as \textit{equilibrium supercurrent nonlinearity}, and the second effect as \textit{non-equilibrium heating nonlinearity}.


In both the BCS theory of superconductivity and the Ginzburg-Landau (GL) theory, the presence of a supercurrent requires the Cooper pairs to have a non-zero superfluid velocity. The equilibrium supercurrent nonlinearity arises as a result of modifications to the superconducting properties in the presence of this superfluid velocity. This nonlinearity can be thought of as an equilibrium effect since the superconducting properties are equilibrium solutions of the BCS/GL theories in the presence of a superfluid velocity term. The supercurrent nonlinearity manifests itself in a variety of measurable properties, such as modified quasiparticle density of states \citenumns{Anthore_2003}, suppressed pair-breaking frequency threshold \citenumns{Zhao_2020}, and suppressed transition temperature \citenumns{songyuan2019_nonlinear}. Importantly to the operation of TWPAs, the kinetic inductance has a dependence on the supercurrent through this effect \citenumns{songyuan2019_nonlinear}. This supercurrent-dependence is often referred to as the nonlinear kinetic inductance or the supercurrent nonlinearity, depending on the emphasis. For the equilibrium supercurrent nonlinearity, in the limit of thin and narrow films, the scaling terms in the kinetic inductance nonlinearity, $I_{*}$ and $I_{*,4}$ in equation~(\ref{eq:nonlinear_scale}), relate to the critical current $I_\mathrm{c}$ by \citenumns{Songyuan_2021_Thesis}
\begin{align}
  I_* &= \frac{3\sqrt{3}}{2} I_\mathrm{c}\,\\
  I_{*,4} &= \frac{1}{\sqrt[4]{3}} I_*\,.
\end{align}

Non-equilibrium heating nonlinearity is another nonlinear effect that emerges in the presence of a strong RF current. Measurements of superconducting resonators have shown that microwave readout power at sub-gap frequencies results in an increased quasiparticle population that contributes to noise and shortens quasiparticle lifetime \citenumns{de_Visser_2010, Goldie_2012, de_Visser_2012, Guruswamy_2018}. 
Microscopically, this observation has been explained by absorption of multiple sub-gap \textit{photons} leading to quasiparticles emitting above-gap \textit{phonons}. The above-gap phonons in turn break Cooper pairs and result in an increased quasiparticle population \citenumns{Goldie_2012}. This pair-breaking interaction results in non-thermal distributions of charge carriers. In this paper, we will henceforth refer to this effect as the non-equilibrium heating nonlinearity. 
The non-equilibrium heating nonlinearity manifests itself strongly in both dissipative and reactive properties of a superconducting transmission line. The kinetic inductance arises through the kinetic energy of the Cooper pairs. In the Drude model, the kinetic inductance is inversely proportional to the pair population \citenumns{Songyuan_2021_Thesis}. By suppressing the Cooper pair population and increasing the quasiparticle population, non-equilibrium heating likewise results in current-dependence in the kinetic inductances.


While these two types of nonlinear phenomena have different origins, a superconductor that conducts a strong RF current will unavoidably display both effects with different magnitudes. It can be difficult to distinguish the two phenomena if one looks at the reactive nonlinearity only: in both theories, the kinetic inductance has the same functional form $L_{\mathrm{k}} = L_{\mathrm{k},0}[1+(I/I_*)^2]$, keeping the lowest order term on current. The sign of the quadratic term is the same for both effects since both nonlinearities increase the kinetic inductance. The two types of nonlinearities can, however, be distinguished by considering their effects on the ratio between the dissipative and reactive nonlinearities and the frequency profile of the wave-mixing processes. These important differences will be discussed in detail in the main body of this paper. 

The relative importance of the two phenomena at a particular readout power level is determined by their onset powers, which is the power required for each nonlinearity to result in the same change in the magnitude of the complex conductivity. For the equilibrium supercurrent nonlinearity, in the limit where the kinetic inductance ratio is close to unity, the onset power can be related straightforwardly to $I_*$ through the characteristic impedance $Z_0$. For the non-equilibrium heating nonlinearity, the onset power can be calculated using the Rothwarf-Taylor equations or more detailed non-equilibrium modelling \citenumns{Rothwarf_1967,Thomas_2020}. When a device is operated in a power regime that is between the onset powers of the two types of nonlinear phenomena, its behaviour will appear to be dominated by the nonlinear mechanism that has the lower onset power. When the operating power is stronger than the onset powers of both mechanisms, the behaviour of a device will be complicated by the interaction between the two mechanisms. This interaction is beyond the scope of this study.

Whilst we have focused on understanding the nonlinearity of a superconductor conceptually in terms of two well-established models, there are other interesting models of nonlinearity published in recent decades \citenumns{Semenov_2016,BELZIG_1999,Larkin_2001,McDonald_thesis,Kubo_2020_2}. There are also other sources of nonlinear behaviour that are driven by current on a superconductor, such as superconducting weak links and local populations of two-level-systems. A detailed discussion of these effects will not be the focus this study. Weak link nonlinearity is present when there are superconducting weak link structures in the superconducting films \citenumns{Likharev_1979}. As yet, these structures have not been reported in the literature to appear in KI-TWPAs. Two-level-system effects, on the other hand, are strongly suppressed at high power levels \citenumns{Gao_2008}. Since TWPAs operate at high power levels in order to achieve efficient nonlinear wave-mixing, it is sufficient to exclude two-level-system nonlinearity at the present level of analysis.

\subsection{Dissipative-reactive ratios}
A general observation in nature is that dissipative and reactive effects are often related, for example through causality \citenumns{Kronig_1926, Kramers_1927}. In this section, we characterise the ratio between the two effects that arise from equilibrium supercurrent nonlinearity and non-equilibrium heating. We have chosen to compare their effect on the complex conductivities $\sigma = \sigma_1-i\sigma_2$, where $\sigma_1$ is the dissipative conductivity and $\sigma_2$ is the reactive conductivity. The effect of the current on $\sigma$ can be straightforwardly inferred from experimental measurements of signal attenuation and phase shift, as shall be shown in the measurements section. In the presence of a strong current, $\sigma$ is modified to be $\sigma+\delta\sigma$, where $\delta\sigma$ characterises the effect that the nonlinear mechanisms have on the complex conductivities, and is given by $\delta\sigma=\delta\sigma_1-i\delta\sigma_2$. We define the dissipative-reactive ratio $\kappa$ to be $\kappa = |\delta\sigma_1/\delta\sigma_2|$.

The effect of the equilibrium nonlinearity on the dissipative-reactive ratio can be calculated using the analysis routine described in \citenumns{songyuan2019_nonlinear, Songyuan_2021_Thesis}. The analysis routine is based on the Usadel equations \citenumns{Usadel1970}, which are reformulated from the BCS theory of superconductivity and are suitable for dirty limit materials or thin-film superconductors \citenumns{Brammertz2004,Kubo_2020}. When the transverse dimensions of the superconducting transmission line are small compared to the superconducting coherence length $\xi_0$, and when then supercurrent is conducted along the length of the transmission line, the Usadel equations have the following form \citenumns{Anthore_2003}:
\begin{align}
E\sin\theta+i\Gamma\cos\theta\sin\theta-i\Delta\cos\theta&=0\,, \\
N_0V\int^{\hbar\omega_{\mathrm{D}}}_0 dE \operatorname{tanh}\left(\frac{E}{2k_\mathrm{B}T}\right)\operatorname{Im}\left(\sin\theta\right)&=\Delta\, ,
\end{align}
where $\theta$ is a complex variable dependent on energy $E$ which parameterises the superconducting properties, $N_\mathrm{0}$ is the single spin density of states at the Fermi surface, $V$ is the interaction potential, $\Delta$ is the superconducting order parameter, $\hbar\omega_{\mathrm{D}}$ is the Debye energy, $T$ is the temperature of the superconducting film, $D$ is the diffusivity constant, given by $D=\sigma_N/(N_0e^2)$ \citenumns{Martinis_2000}, $e$ is the elementary charge, $\operatorname{Im}(x)$ takes the imaginary part of $x$, and finally $\sigma_N$ is the normal state conductivity, at $T$ just above the superconducting transition temperature $T_c$. The effect of the current is encapsulated using the depairing term $\Gamma$.

Although not needed for this study, solutions of superconducting properties in term of $\Gamma$ can be expressed in terms of supercurrent density $\vec{j}$ by:
\begin{align}
  \Gamma &= \frac{\hbar}{2D_S}\vec{v}_s^2\, \\
  \vec{j} &= q_\mathrm{cp} N_{\mathrm{N,cp}} \vec{v}_s\,, \\
  N_{\mathrm{N,cp}} &= \frac{1}{2} N_0 \int_0^{\infty} \tanh{\left(\frac{E}{2k_\mathrm{B}T}\right)}\mathrm{Im}(\sin^2\theta)\,,
\end{align}
where $q_\mathrm{cp}=2e$ is the charge of a Cooper pair, $\vec{v}_s$ is the superfluid velocity, and $N_{\mathrm{N,cp}}$ is the number density of Cooper pairs. At absolute zero temperature, the critical current density is reached when $\Gamma=\Delta_0/4$, where $\Delta_0$ is the density of states gap at absolute zero, given in the BCS theory by $\Delta_0=1.764\,k_\mathrm{B}T_\mathrm{c}$, where $k_\mathrm{B}$ is the Boltzmann constant. Thus the range of physically meaningful $\Gamma/\Delta_0$ lies between $0$ and $0.25$ for this study.

Numerically, the Usadel equations were used to calculate the effect of $\Gamma$ on the densities of states $\cos\theta$ and $\sin\theta$. Afterwards, the densities of states were integrated using Nam's equations \citenumns{Nam_1967, Songyuan_2018} to obtain the complex conductivities.

\begin{figure}[htbp]
  \centering
\includegraphics[width=8.6cm]{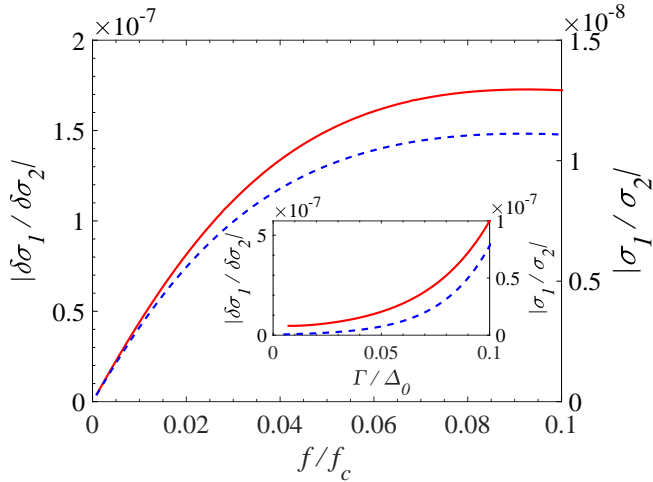}
\caption{\label{fig:Intrinsic_DR} Main figure: solid red line, left axis: dissipative-reactive ratio $\kappa=|\delta \sigma_1 / \delta \sigma_2|$ arising from the equilibrium nonlinearity plotted against scaled frequency $f/f_c$; dashed blue line, right axis: ratio of conductivities $\sigma_1/\sigma_2$ arising from the equilibrium nonlinearity plotted against frequency $f/f_c$. Calculation is performed for Al at temperature $T=0.1\,\mathrm{K}$ and depairing term $\Gamma/\Delta_0=0.05$. \\
Inset: solid red line, left axis: dissipative-reactive ratio $\kappa=|\delta \sigma_1 / \delta \sigma_2|$ arising from the equilibrium nonlinearity plotted against scaled depairing term $\Gamma/\Delta_0$; dashed blue line, right axis: ratio of conductivities $\sigma_1/\sigma_2$ arising from the equilibrium nonlinearity plotted against scaled depairing term $\Gamma/\Delta_0$. Calculation is performed for Al at temperature $T=0.1\,\mathrm{K}$ and frequency $f=3\,\mathrm{GHz}$.}
\end{figure}

The dissipative-reactive ratio $\kappa$ arising from the equilibrium nonlinearity is shown in figure~\ref{fig:Intrinsic_DR} for Al at $T=0.1\,\mathrm{K}$. The main figure shows the dependence of $\kappa$ on the scaled frequency $f/f_c$ whereas the inset shows the dependence of $\kappa$ on the scaled depairing term $\Gamma/\Delta_0$. The scaling frequency $f_c$ is the pair-breaking frequency of Al when $T\ll T_c$, given by $f_c=2\Delta_0 / h$, $h$ is the Planck constant. The dissipative part of the equilibrium supercurrent nonlinearity remains very small compared to the reactive part when the supercurrent frequency is small compared to the pair-breaking frequency of the material, which is $\sim88\,\mathrm{GHz}$ for Al. On a travelling-wave transmission line, this small dissipative change can be extremely hard to measure because it is often dominated by other dissipation mechanisms such as dielectric loss. It can instead be observed experimentally by measuring the change in quality factor of a high-Q superconducting resonator in response to a bias current \citenumns{Vissers_2015}.

The effect of non-equilibrium heating can be calculated by considering the change in $\sigma_1$ and $\sigma_2$ in response to a change in quasiparticle number density due to pair-breaking. This topic has been well studied in the context of kinetic inductance detectors and the results are as follows \citenumns{Jonas_review}:
\begin{align}
\frac{\delta\sigma_{1,2}}{\sigma_N}&=\frac{\pi}{2N_0\hbar\omega}S_{1,2}(\omega,T,\Delta_0)\delta n_{\mathrm{qp}} \label{eq:QP_Heating}\\
S_{1}(\omega,T,\Delta_0) &= \notag \\ &\frac{2}{\pi}\sqrt{\frac{2\Delta_0}{\pi k_{B}T}}\mathrm{sinh}\left(\frac{\hbar\omega}{2k_{B}T}\right) \mathrm{K_0}\left(\frac{\hbar\omega}{2k_{B}T}\right) \label{eq:QP_Heating_1}\\
S_{2}(\omega,T,\Delta_0) &= \notag \\ 1&+\sqrt{\frac{2\Delta_0}{\pi k_{B}T}}\mathrm{exp}\left(\frac{-\hbar\omega}{2k_{B}T}\right) \mathrm{I_0}\left(\frac{\hbar\omega}{2k_{B}T}\right)\,,\label{eq:QP_Heating_2}
\end{align}
where $\mathrm{I_0}$ and $\mathrm{K_0}$ are the modified Bessel functions of the first and second kind respectively, $\hbar$ is the reduced Planck constant, $\omega$ is the angular frequency, and $\delta n_{\mathrm{qp}}$ is the change in quasiparticle number density. Equation~(\ref{eq:QP_Heating}) gives the dissipative-reactive ratio to be $\kappa = |S_{1}/S_{2}|$. For BCS superconductors, both $\Delta_0$ and $f_c$ scale linearly with $T_c$ according to $\Delta_0=1.764\,k_\mathrm{B}T_\mathrm{c}$ and $f_c = 2\Delta_0/h$. As a result, equations~(\ref{eq:QP_Heating_1}) and (\ref{eq:QP_Heating_2}) can be expressed generally in terms of scaled frequency $f/f_c$ and scaled temperature $T/T_c$ as follows:
\begin{align}
S_{1}\left(\frac{f}{f_c},\frac{T}{T_c}\right) &= \notag \\
&\frac{2}{\pi}\sqrt{\frac{2\alpha}{\pi}\frac{T_c}{T}}\mathrm{sinh}\left(\alpha\frac{f}{f_c}\frac{T_c}{T}\right) \mathrm{K_0}\left(\alpha\frac{f}{f_c}\frac{T_c}{T}\right) \label{eq:QP_Heating_3}\\
S_{2}\left(\frac{f}{f_c},\frac{T}{T_c}\right)
&= \notag \\ 1&+\sqrt{\frac{2\alpha}{\pi}\frac{T_c}{T}}\mathrm{exp}\left(-\alpha\frac{f}{f_c}\frac{T_c}{T}\right) \mathrm{I_0}\left(\alpha\frac{f}{f_c}\frac{T_c}{T}\right)\,,\label{eq:QP_Heating_4}
\end{align}
where $\alpha = 1.764$.

\begin{figure}[htbp]
  \centering
\includegraphics[width=8.6cm]{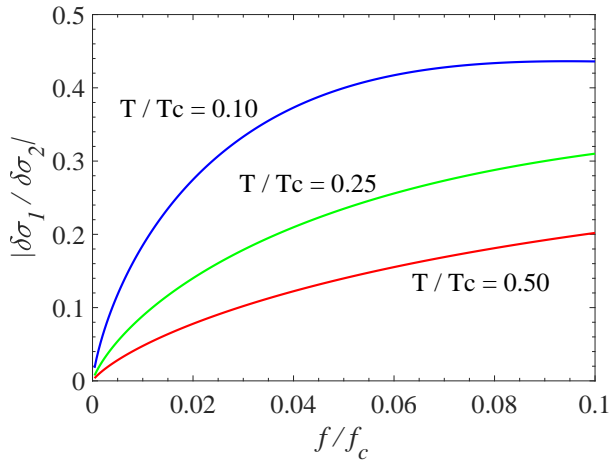}
\caption{\label{fig:QP_Heating_DR} Dissipative-reactive ratio $\kappa=|\delta \sigma_1 / \delta \sigma_2|$ arising from non-equilibrium heating nonlinearity plotted against scaled frequency $f/f_c$. Red lines shows the behaviour at scaled tempearture $T/T_c=0.50$; green lines shows the behaviour at $T/T_c=0.25$; blue lines shows the behaviour at $T/T_c=0.10$.}
\end{figure}

Figure~\ref{fig:QP_Heating_DR} shows the dissipative-reactive ratio $\kappa$ for BCS superconductors arising from non-equilibrium heating nonlinearity plotted against scaled frequency $f/f_c$ at scaled temperatures $T/T_c=0.50,\,0.25,\,0.10$. The values for the reduced temperature have been chosen to reflect those for a Nb film with $T_c=8.0\,\mathrm{K}$ cooled by a pulse tube cooler ($4\,\mathrm{K}$), and an Al film with $T_c=1.2\,\mathrm{K}$ cooled by a He-3 adsorption cooler ($300\,\mathrm{mK}$) and an adiabatic demagnetization refrigerator ($120\,\mathrm{mK}$). Comparing figure~\ref{fig:Intrinsic_DR} and figure~\ref{fig:QP_Heating_DR}, we see that $\kappa$ arising from non-equilibrium heating nonlinearity is significantly greater. As figure~\ref{fig:QP_Heating_DR} shows, when we vary the frequency of the power sweeps, the $\kappa$ arising from non-equilibrium heating nonlinearity increases in magnitude until it reaches a maximum. In comparison, $\kappa$ arising from the equilibrium supercurrent nonlinearity remains close to zero regardless of frequency, provided that the sweep frequency is far from the pair-breaking frequency.

\subsection{Response time model of nonlinear inductance}
The model of nonlinearity described by equation~(\ref{eq:nonlinear_scale}) implicitly assumes that the rate at which the equivalent circuit element responds is instantaneous. Our response time model instead characterises the nonlinear mechanism by a finite response time $\tau$. The value of $\tau$ depends on the dominant physical mechanism responsible for the nonlinearity. For example, the non-equilibrium heating nonlinearity can be characterised by the quasiparticle recombination lifetime \citenumns{Thomas_2020,Jonas_review} which can be controlled by trapping and diffusion \citenumns{Goldie_1990,Loidl_2001,Wang2014}; likewise, the equilibrium supercurrent nonlinearity can be characterised by a relaxation time \citenumns{Lucas_Relaxation_1967, Mattoo_1982effect, Stuivinga_1983, Mattoo_1983}. In the response time model, the inductance per unit length now takes a modified form:
\begin{align}
  L&=L_0+L_{\mathrm{NL}}\,, \label{eq:Response_0}\\
  L_{\mathrm{NL}}(t)&=\frac{L_0}{I_*^2}\int_{-\infty}^{t} \frac{1}{\tau}\exp\left(-\frac{t-t'}{\tau}\right)I^2(t') dt' \,, \label{eq:Response}
\end{align}
where $L$ is the total inductance, $L_0$ is the unperturbed inductance in the absence of nonlinearity, $L_{\mathrm{NL}}$ is the nonlinear contribution to inductance, and $t$ is time. The mathematical form of equation~(\ref{eq:Response}) is motivated by the fact that the nonlinear inductance is a kinetic energy phenomenon and is related to even orders of the current. The response time model gives the simplest time dependence to the energy in the system that captures an exponential decay characterised by a timescale $\tau$. In the case of instantaneous response, i.e. $\tau\rightarrow0$, equations~(\ref{eq:Response_0}) and (\ref{eq:Response}) give
\begin{align}
  L(t) \rightarrow L_0\left[1+\left(\frac{I(t)}{I_*}\right)^2\right]=L_0+L_{\mathrm{NL},0}(t)\,.
\end{align}
Thus in the limit where the nonlinear mechanism is instantaneous, we recover the familiar form of inductance nonlinearity. Here $L_{\mathrm{NL},0}(t)$ is the instantaneous nonlinear inductance.


\begin{figure}[htbp]
  \centering
\includegraphics[width=8.6cm]{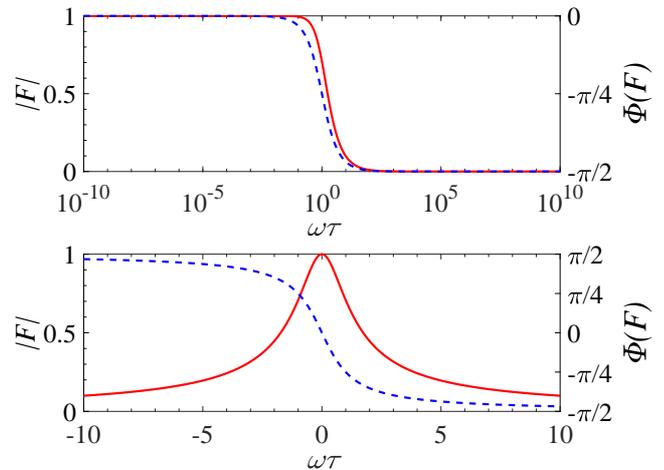}
\caption{\label{fig:filter_comb} Top figure: solid red line: magnitude of response time filter $|F|$ plotted against angular frequency scaled by the response time $\omega\tau$; dashed blue line: phase of response time filter $\Phi(F)$ plotted against frequency scaled by the response time $\omega\tau$. The horizontal axis is \textit{logarithmic}.\\
Bottom figure: solid red line: magnitude of response time filter $|F|$ plotted against angular frequency scaled by the response time $\omega\tau$; dashed blue line: phase of response time filter $\Phi(F)$ plotted against frequency scaled by the response time $\omega\tau$. The horizontal axis is \textit{linear}.}
\end{figure}

It is convenient to work in frequency domain. The Fourier-transforms of the time-dependent inductances are defined by
\begin{align}
  L_{\mathrm{NL}}(\omega)&=\int_{-\infty}^{\infty}L_{\mathrm{NL}}(t)\exp(-i\omega t) dt\,, \label{eq:Fourier} \\
  L_{\mathrm{NL,0}}(\omega)&=\int_{-\infty}^{\infty}L_{\mathrm{NL,0}}(t)\exp(-i\omega t) dt\,,
\end{align}
where $\omega$ is the angular frequency. Substituting equation~(\ref{eq:Response}) into equation~(\ref{eq:Fourier}), it can be shown that
\begin{align}
  L_{\mathrm{NL}}(\omega)&=F(\omega)L_{\mathrm{NL,0}}(\omega)\,, \label{eq:filter} \\
  F(\omega)&=\frac{1}{1+i\omega\tau}\,,
\end{align}
where $F(\omega)$ can be interpreted as a frequency filter applied to the nonlinear inductance as a result of the response time of the underlying microscopic processes. It is important to note that the filter is applied to perturbations in inductance and not directly to current or voltage: it places limitations on the \textit{bandwidth} of wave-mixing instead of the maximum frequency of \textit{transmission}.


Figure~\ref{fig:filter_comb} shows the amplitude and phase of $F$ as a function of normalised frequency $\omega\tau$. In the top panel the frequency axis is \textit{logarithmic}, whereas in the bottom panel it is \textit{linear} and extends over positive and negative frequencies. As expected, $F$ shows low-pass behaviour with the transmission magnitude decreasing rapidly for $\omega\tau > 1$. This will result in a limitation on the amplification bandwidth. As seen in bottom panel, the imaginary part of the filter dominates at higher frequencies and results in a $\pi$ phase shift when crossing from $\omega\ll-1/\tau$ to $\omega\gg 1/\tau$. This phase shift results in a characteristic feature in the amplifier gain profile, which we will describe shortly.  

The filter $F(\omega)$ can be straightforwardly applied to coupled mode numerical analyses of wave-mixing in TWPAs, such as the models presented in \citenumns{Songyuan_2019_paramp}, chapter 5 of \citenumns{Songyuan_2021_Thesis}, and the appendix of \citenumns{Eom_2012}. Coupled mode solvers naturally handle quantities decomposed into their frequency components. The response time frequency filter can thus be applied multiplicatively to the frequency components of the inductances directly. Here we illustrate the process of applying the response time model to the coupled mode analysis of a four-wave-mixing TWPA.

In general, current on a transmission line $I$ can be decomposed into its modes of propagation $I_{\mathrm{n}}$ such that
\begin{align}
  &I = \sum_{\mathrm{n}} I_{\mathrm{n}}\,,\label{eq:CM_sum}\\
  &I_{\mathrm{n}} =\frac{1}{2}A_{\mathrm{n}}(z)\operatorname{exp}(i\omega_{\mathrm{n}} t-\gamma_{\mathrm{n}} z)\,, \label{eq:CM_prop}\\
  &I_{-\mathrm{n}} = I_{\mathrm{n}}^* \,, \label{eq:CM_conj}
\end{align}
where suffix $n$ labels each propagating mode, $\omega_{\mathrm{n}}$ is the angular frequency, $\gamma_{\mathrm{n}}$ is the complex propagation constant in the absence of inductive nonlinearity, and $A_{\mathrm{n}}(z)$ is the position varying amplitude of the current mode. Here the exponential term contains the propagation behaviour of a linear transmission line, and the $A_{\mathrm{n}}(z)$ term contains the effect of the nonlinearity. Equation~(\ref{eq:Response}) can be expressed in terms of these propagating modes as follows
\begin{align}
  L_{\mathrm{NL}}(t)&=\sum_{\mathrm{i},\mathrm{j}}L_{\mathrm{NL}}(I_\mathrm{i},I_\mathrm{j},t) \\
  L_{\mathrm{NL}}(I_\mathrm{i},I_\mathrm{j},t)&=\frac{L_0}{I_*^2}\int_{-\infty}^{t} \frac{1}{\tau}\exp\left(-\frac{t-t'}{\tau}\right)I_\mathrm{i}(t')I_\mathrm{j}(t') dt' \,.
\end{align}
The transmission line model of a TWPA can be analysed using the telegrapher's equations to give
\begin{align}
 \pdv[2]{I}{z} &=\pdv{}{t}\left[LC\pdv{I}{t}\right] \,\\
 &=L_0C\pdv[2]{I}{t}+\pdv{}{t}\left[L_{\mathrm{NL}}C\pdv{I}{t}\right].
\end{align}
We shall refer to the first term as the linear propagation term, and the second term as the nonlinear mixing term.
The above equation can be expressed in term of the propagating modes, giving
\begin{align}
\pdv[2]{I_\mathrm{n}}{z} =& L_0C\pdv[2]{I_\mathrm{n}}{t} \notag \\
&+\sum_{\omega_\mathrm{i}+\omega_\mathrm{j}+\omega_\mathrm{k}=\omega_\mathrm{n}}\pdv{}{t}\left[L_{\mathrm{NL}}(I_\mathrm{i},I_\mathrm{j},t)C\pdv{I_\mathrm{k}}{t}\right]\,.
\end{align}
The response time frequency filter can be applied by noting
\begin{align}
L_{\mathrm{NL}} (I_\mathrm{i},I_\mathrm{j},t) = F(\omega_\mathrm{i}+\omega_\mathrm{j})L_{\mathrm{NL},0} (I_\mathrm{i},I_\mathrm{j},t)\,.
\end{align}
In summary, the coupled differential equations governing the evolution of the propagating modes are modified by applying the response time frequency filter multiplicatively to the nonlinear mixing term. At this stage, the system of equations can be simplified and solved using the usual routine \citenumns{Songyuan_2021_Thesis}: cancelling the linear propagation term, applying the slowly-varying-envelope-approximation, and numerically solving the simplified differential equations. In this study, we calculate the gain profiles by applying the response time model to the coupled mode solver described in chapter 5 of \citenumns{Songyuan_2021_Thesis}, which restricts the analysis to pump, signal, and idler modes.


\begin{figure}[htbp]
  \centering
\includegraphics[width=8.6cm]{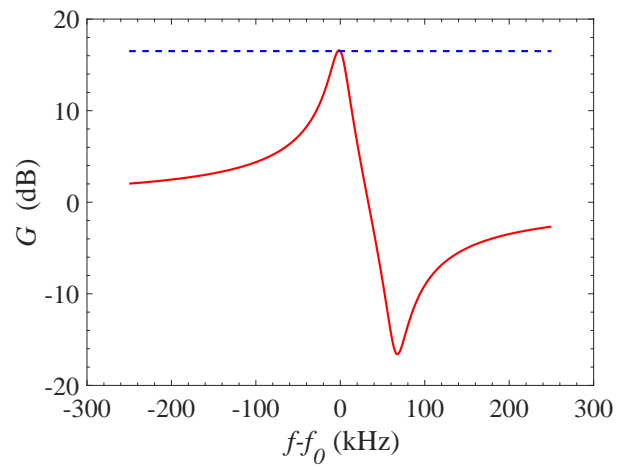}
\caption{\label{fig:slow_NL_Gain} Power gain $G$ of a Ti TPWA plotted against $f-f_0$ which is the difference between measurement frequency $f$ and pump frequency $f_0 = 10\,\mathrm{GHz}$. The TWPA has length $l=0.5\,\mathrm{m}$ and is operated with a pump current of amplitude $I_{p,0}=0.15\,I_*$, where $I_*$ is the characteristic nonlinear scale of current. Solid red line: the underlying nonlinearity has response time of $\tau=1/(2\pi\times10^{4})\,\mathrm{s}$; dashed blue line: the underlying nonlinear response is instantaneous.}
\end{figure}

Figure~\ref{fig:slow_NL_Gain} shows the simulated gain profile of a $0.5\,\mathrm{m}$ long Ti TWPA. The solid red line assumes $\tau=1/(2\pi\times10^{4})\,\mathrm{s}$ whereas the dash blue line assumes that the underlying nonlinear response is instantaneous. As seen in the figure, the response time filter restricts the bandwidth of the gain profile. This bandwidth restriction can be illustrated by considering the frequency mixing modes that contribute to signal amplification. In a 4-wave-mixing system, \textit{one of} the mixing modes is as follows: an idler at $\omega_i$ and a pump at $\omega_p$ generates their difference frequency $\Delta\omega=\omega_p-\omega_i=\omega_s-\omega_p$ in the inductance. This difference frequency simultaneously mixes with $\omega_p$ to generate amplified signal $\omega_p+\Delta\omega=\omega_s$. In total, two pump modes are coupled with one idler mode to generate one signal mode: this mixing mode is thus allowed by 4-wave-mixing. 
The slow response time filtering however, attenuates the amplitude of $L_{\mathrm{NL}}$ where $\Delta\omega>1/\tau$. This leads to smaller gain at larger $\Delta\omega$. In the gain profile this effect restricts the bandwidth to the order of $1/\tau$. The other mixing modes can be shown to be similarly affected by the inductance response time filtering.


Another characteristic feature demonstrated by figure~\ref{fig:slow_NL_Gain} is the gain dip at $f-f_0\sim60\,\mathrm{kHz}$. This effect can be attributed to the $\pi$ phase flipping of $F(\omega)$ when the angular frequency crosses from $\omega\ll-1/\tau$ to $\omega\gg 1/\tau$. The output signal consists of the original signal $I_{\mathrm{s,0}}$ and the signal generated by nonlinear inductance $I_{\mathrm{s,NL}}$. As the two contributions are added together, the phase flipping at signal frequencies higher than the pump frequency induces a destructive interference effect. This interference effect is hidden when $I_{\mathrm{s,NL}}\ll I_{\mathrm{s,0}}$ or $I_{\mathrm{s,NL}}\gg I_{\mathrm{s,0}}$. It is obvious on the gain profile when $I_{\mathrm{s,NL}}\sim I_{\mathrm{s,0}}$. If, in simulations, a new filter is imposed such that $F_{\mathrm{new}}(\omega)=|F(\omega)|$, the dip in gain disappears. This confirms the gain dip is related to the phase-shift introduced by the filtering.

\section{Measurements of Al and Ti travelling-wave devices}
The results in this section were obtained from measurements of superconducting Al and Ti TWPA devices at a temperature of $0.1\,\mathrm{K}$. 
Al and Ti were chosen because calculations based on the BCS theory have shown that these materials have low onset power with respect to the equilibrium supercurrent nonlinearity \citenumns{songyuan2019_nonlinear, Songyuan_2021_Thesis}. The devices presented in this work were based on co-planar waveguide (CPW) geometries. The centre conductor widths were $20\, \mathrm{\mu m}$ and the gap widths were $2\, \mathrm{\mu m}$ for all devices. The Al devices had thicknesses of $50\,\mathrm{nm}$ whereas the Ti devices had thicknesses of $75\,\mathrm{nm}$. These devices were designed to have close to $50\,\mathrm{\Omega}$ characteristic impedance and high kinetic inductance ratios. Following the work by Eom  \textit{et al.} \citenumns{Eom_2012}, periodic impedance loading features as described by \citenumns{Songyuan_2019_paramp} had been included to improve gain.

Films were deposited onto 2-inch SEMI standard Si wafers by DC magnetron sputtering at a base pressure of $2\times 10^{-10}\,\,\rm{Torr}$ or below. Ti films were deposited at ambient temperature and Al films were deposited after substrate cooling to liquid nitrogen temperatures. The films were patterned via a lift-off process using an image-reversal mode photoresist. The samples were attached to the cold stage of an adiabatic demagnetization refrigerator and the temperature was monitored using a calibrated ruthenium oxide thermometer. The measurements of power transmission and phase shift were obtained using a vector network analyser (VNA). In this study, the power transmission $T_\mathrm{out}$ is defined as the ratio between the power received at the VNA $P_\mathrm{out}$ and the power delivered by the VNA into the cryogenic system $P_\mathrm{in}$. It includes both the effect of the TWPA under test and the attenuation due to coaxial cables in the cryostat, which was about $55 \,\mathrm{dB}$ at $8.0\,\mathrm{GHz}$. When all quantities are measured in units of \textrm{dB}, $T_\mathrm{out}$ is given by $T_\mathrm{out}=P_\mathrm{out}-P_\mathrm{in}$.

\subsection{Dissipative-reactive ratios}
The dissipative-reactive ratio of a device can be inferred from measurements of its nonlinear power attenuation and nonlinear phase shift in response to increasing readout power. Nonlinear power attenuation $T_\mathrm{\alpha}$ is a function of the input power $P_\mathrm{in}$, and is defined as $T_\mathrm{\alpha}(P_\mathrm{in})=T_\mathrm{out,0}-T_\mathrm{out}(P_\mathrm{in})$, where $T_\mathrm{out,0}$ is the power transmission ratio at very low input power below the onset of nonlinearity. The phase shift $\Delta\Phi$ gives the phase difference between the output signal and the input signal. The nonlinear phase shift $\Delta\Phi_\mathrm{NL}$ is given by $\Delta\Phi_\mathrm{NL}(P_\mathrm{in})=\Delta\Phi_{0}-\Delta\Phi(P_\mathrm{in})$, where $\Delta\Phi_{0}$ is the phase shift at very low input power below the onset of nonlinearity.

From transmission line theory, for small perturbations, it can be shown that \citenumns{Songyuan_transmission_lines_2018,Songyuan_2021_Thesis}
\begin{align}
T_\mathrm{\alpha} &= 8.69\, L\,\delta\alpha\,\,\,\,\,\,\,\mathrm{dB},\\
\Delta\Phi_\mathrm{NL} &=|L\,\delta\beta|\,,
\end{align}
where $L$ is the length of the transmission line, $\delta\alpha$ is the change in attenuation constant, $\delta\beta$ is the change in phase constant. $\delta\alpha$ and $\delta\beta$ are related to the change in full propagation constant $\delta\gamma$ by $\delta\alpha+i \delta\beta=\delta \gamma$. Further, for a small input power, it can be shown that
\begin{align}
  \delta\gamma &= C_{\mathrm{geom}}\delta\sigma \\
  \kappa &= \left|\frac{\delta\sigma_1}{\delta\sigma_2}\right| = \left|\frac{\delta\alpha}{\delta\beta}\right|\,, \label{eq:DR_ratio}
\end{align}
where $C_{\mathrm{geom}}$ is a constant of proportionality arising from device geometry. Thus $\kappa$ for a particular device can be deduced by relating its nonlinear attenuation $T_\mathrm{\alpha}$ to its nonlinear phase shift $\Delta\Phi_\mathrm{NL}$.


\begin{figure}[htbp]
  \centering
\includegraphics[width=8.6cm]{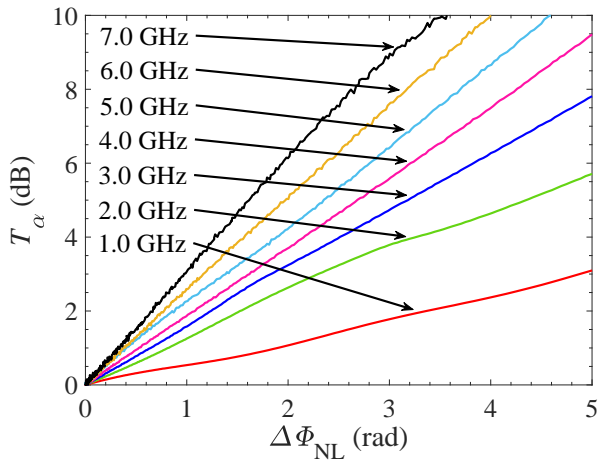}
\caption{\label{fig:Ti_CPW_01_collated}  Nonlinear power attenuation $T_\mathrm{\alpha}$ against nonlinear phase shift $\Delta\Phi_\mathrm{NL}$ for a TWPA based on $50\,\mathrm{cm}$ long, $75\,\mathrm{nm}$ thick Ti CPW transmission line. Frequencies of the underlying power sweep are $1.0\,\mathrm{GHz},\,2.0\,\mathrm{GHz},\,3.0\,\mathrm{GHz},\,4.0\,\mathrm{GHz},\,5.0\,\mathrm{GHz},\,6.0\,\mathrm{GHz}$, and $7.0\,\mathrm{GHz}$.}
\end{figure}

Figure~\ref{fig:Ti_CPW_01_collated} shows the nonlinear power attenuation $T_\mathrm{\alpha}$ against nonlinear phase shift $\Delta\Phi_\mathrm{NL}$ for a Ti CPW device at frequencies $1.0\,\mathrm{GHz},\,2.0\,\mathrm{GHz},\,3.0\,\mathrm{GHz},\,4.0\,\mathrm{GHz},\,5.0\,\mathrm{GHz},\,6.0\,\mathrm{GHz}$, and $7.0\,\mathrm{GHz}$. As seen in the plot, as frequency increases, so does the relative magnitude of the dissipative nonlinearity.

\begin{figure}[htbp]
  \centering
\includegraphics[width=8.6cm]{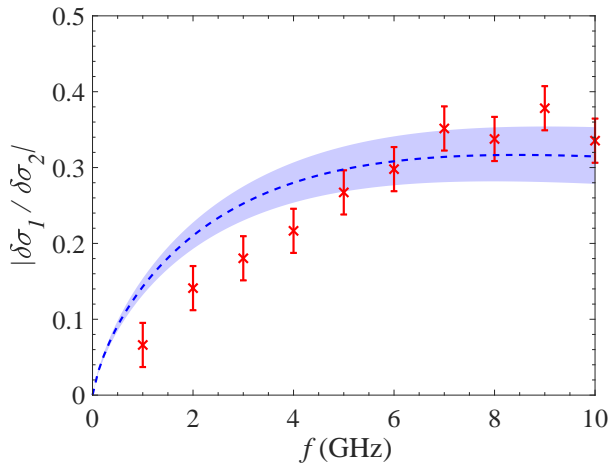}
\caption{\label{fig:Ti_CPW_01_fitted}  Dissipative-reactive ratio $\kappa=|\delta \sigma_1 / \delta \sigma_2|$ plotted against frequency $f$. The red points show data for a Ti CPW transmission line, as obtained by fitting the gradients of the lines in figure \ref{fig:Ti_CPW_01_collated}; the blue dashed line shows the best fit using the non-equilibrium heating model. Transition temperature $T_c$ is the fit variable and experimentally recorded bath temperature $T=0.15\,\mathrm{K}$ is a fixed input to the model. The best fit transition temperature is $0.44\,\mathrm{K}$ with a $90\%$ confidence interval of $0.30\,\mathrm{K}$ to $0.66\,\mathrm{K}$. The range of behaviour described by the $90\%$ confidence interval is shown by the light blue region surrounding the best fit line.}
\end{figure}

Figure~\ref{fig:Ti_CPW_01_fitted} shows the experimentally measured $\kappa$ against frequency $f$. The red points show data for the Ti CPW device described previously, as obtained by fitting the gradients of the lines in figure \ref{fig:Ti_CPW_01_collated}. The blue dashed line shows the least squares fit using the non-equilibrium heating model. The bath temperature was fixed at the experimentally measured value $T=0.15\,\mathrm{K}$, leaving the transition temperature $T_c$ as the only fitting variable. We have assumed that Ti is a BCS superconductor and can be described by $\Delta_0=1.764\,k_\mathrm{B}T_\mathrm{c}$.  The best fit transition temperature is $0.44\,\mathrm{K}$ with a $90\%$ confidence interval of $0.30\,\mathrm{K}$ to $0.66\,\mathrm{K}$. The range of behaviour described by the $90\%$ confidence interval is shown by the light blue region surrounding the best fit line. As seen in the figure, the magnitude and frequency trend of $\kappa$ are similar to those predicted by non-equilibrium heating. Small deviations to the model may be the result of 1) on-chip heating due to measurements performed at high powers \citenumns{Songyuan_2021_Thesis}, 2) a small contribution from the equilibrium supercurrent nonlinearity increasing the measured reactive nonlinearity, and 3) spurious ground-plane resonances due to the large size of the chip \citenumns{Adamyan_2016}. The variations occurring in the range $6\,\mathrm{GHz}$ to $10\,\mathrm{GHz}$ can be used as an estimate of the magnitude of uncertainties and secondary effects, as shown in the error bars of figure~\ref{fig:Ti_CPW_01_fitted}. The best fit transition temperature of $0.44\,\mathrm{K}$ is consistent with direct measurements of the transition temperatures of Ti in the literature, which span the range of $0.37\,\mathrm{K}$ to $0.59\,\mathrm{K}$ \citenumns{Ti_Steele_1953, Zhao_2018}. Overall, the behaviour of the Ti transmission line is well described by the model, suggesting that the non-equilibrium heating mechanism contributes strongly to the nonlinearity displayed by the Ti device.

\subsection{Gain profiles}
In this study, the power gain $G$ of a device is defined as the ratio between the output signal power transmission in the presence of a strong pump current $T_\mathrm{out,P}$ generated by a signal generator and the output signal power transmission in the absence of a strong pump current $T_\mathrm{out,0}$. Since both $T_\mathrm{out,P}$ and $T_\mathrm{out,0}$ are measured in units of $\mathrm{dB}$, the power gain is given by $G=T_\mathrm{out,P}-T_\mathrm{out,0}$.

\begin{figure}[htbp]
  \centering
\includegraphics[width=8.6cm]{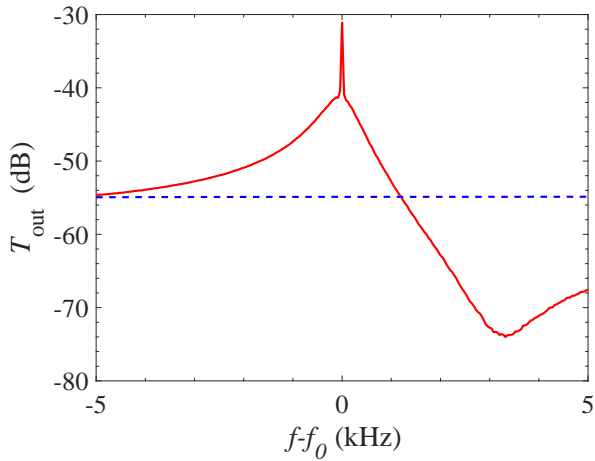}
\caption{\label{fig:Al_gain_measurement_Wide}  Power transmission measurement of an Al CPW TWPA. The horizontal axis indicates the difference between measurement frequency $f$ and pump frequency $f_0 = 8.0\,\mathrm{GHz}$. Pump power from at the output of signal generator is $14.66\,\mathrm{dBm}$. Solid red line: power transmission in the presence of pump current; dashed blue line: power transmission in the absence of pump current. The blue line characterises the losses in the test system. The difference between the red and the blue line indicates device gain.}
\end{figure}

Figure~\ref{fig:Al_gain_measurement_Wide} shows the power transmission measurement of a CPW TWPA based on Al. The TWPA transmission line is biased at pump frequency $f_0 = 8.0\,\mathrm{GHz}$ with a pump power of $14.66\,\mathrm{dBm}$, as measured at the output of the VNA. The solid red line shows power transmission in the presence of pump current whereas the dashed blue line shows power transmission in the absence of pump current. The transmission spike at $f-f_0=0\,\mathrm{kHz}$ is the unsuppressed pump current. As seen in the figure, without the biasing pump current, the transmission of the TWPA line is at $-55\,\mathrm{dB}$. This attenuation is mainly a result of losses from long coaxial cables which are external to the chip. Once the pump has been switched on, the signal is amplified to as high as $-41\,\mathrm{dB}$. Thus a maximum on-chip gain of $14\,\mathrm{dB}$ is obtained. Interestingly, the $3\,\mathrm{dB}$ bandwidth of the device is only around $1\,\mathrm{kHz}$, instead of in the range of $\mathrm{GHz}$. The observation of this limited bandwidth strongly motivated the formulation of the response time model of nonlinear inductance that we have presented.

\begin{figure}[htbp]
  \centering
\includegraphics[width=8.6cm]{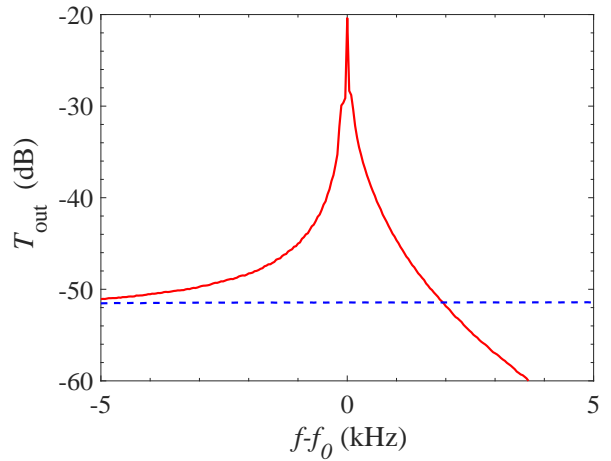}
\caption{\label{fig:Al_gain_measurement}  Power transmission measurement of an Al CPW TWPA, optimised for gain. The horizontal axis indicates the difference between measurement frequency $f$ and pump frequency $f_0 = 7.0\,\mathrm{GHz}$. Pump power from at the output of signal generator is $13.80\,\mathrm{dBm}$. Solid red line: power transmission in the presence of pump current; dashed blue line: power transmission in the absence of pump current. The blue line characterises the losses in the test system. The difference between the red and the blue line indicates device gain.}
\end{figure}
Figure~\ref{fig:Al_gain_measurement} shows the power transmission measurement of the same device with an optimised pump power of $13.80\,\mathrm{dBm}$ at $f_0 = 7.0\,\mathrm{GHz}$. Optimisation is necessary because the devices show strong dissipative nonlinearity at high pump powers. This causes the gain to increase, peak, and then decrease again as the pump power is increased. This optimisation is performed by slowly varying the magnitude of the pump power at each frequency to obtain the largest gain. 
As figure~\ref{fig:Al_gain_measurement} demonstrates, by optimising the pump current, we are able to obtain peak gain of more than $20\,\mathrm{dB}$.

\begin{figure}[htbp]
  \centering
\includegraphics[width=8.6cm]{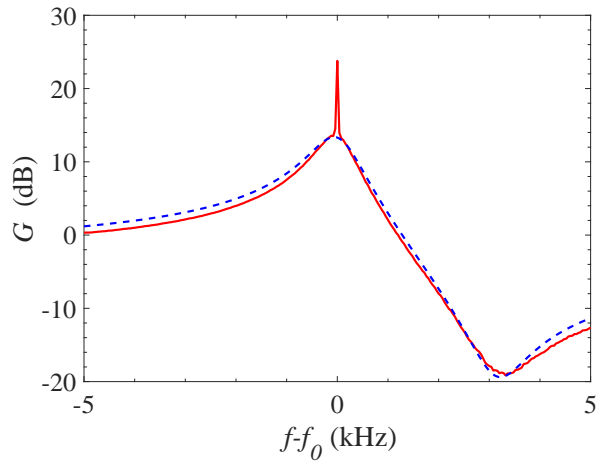}
\caption{\label{fig:gain_Data_fitted} Gain data of an Al CPW TWPA. The horizontal axis indicates the difference between measurement frequency $f$ and pump frequency $f_0 = 8.0\,\mathrm{GHz}$. Pump power from at the output of signal generator is $14.66\,\mathrm{dBm}$. Solid red line: gain data obtained experimentally; dashed blue line: gain data as calculated using response time nonlinearity model. The best fit $\tau$ of $3.27\times10^{-4}\,\mathrm{s}$ is used.}
\end{figure}

Figure~\ref{fig:gain_Data_fitted} shows a comparison between the gain obtained experimentally and the gain calculated using the response time nonlinearity model. We have included $3\,\mathrm{dB}$ of best fit dissipative loss in this calculation to account for the fact that the presence of a strong pump also introduces nonlinear dissipation, as discussed in earlier sections. The best fit $\tau$ of $3.27\times10^{-4}\,\mathrm{s}$ is used. As seen in the figure, the model describes experimental data very well and captures the following key characteristic behaviour: 1) height of gain around pump frequency; 2) $3\,\mathrm{dB}$ bandwidth; 3) the gain dip at $\sim3.2\,\mathrm{kHz}$; and 4) the gradients of the rise and fall in gain around the maximum. The measured timescale of $3.27\times10^{-4}\,\mathrm{s}$ is much shorter than the timescale of the equilibrium superconductor nonlinearity, which should be in the range of nanoseconds to picoseconds \citenumns{Lucas_Relaxation_1967, Mattoo_1982effect, Stuivinga_1983, Mattoo_1983}, but is similar to the timescale of quasiparticle recombination, which should be in the range of milliseconds to microseconds \citenumns{Barends_Lifetime_2008, deVisser2012, Jonas_review}. This suggests that the non-equilibrium heating mechanism contributes strongly to the nonlinearity displayed by the Al device.

\section{Conclusions}
We have described a method of characterising the underlying nonlinearity of a superconducting material by its dissipative-reactive ratio and the response time of its nonlinear mechanism. Both quantities are important in determining the performance of TWPAs, such as in terms of added noise, maximum gain, and bandwidth. We have discussed how different nonlinearities, for example the equilibrium supercurrent nonlinearity and the non-equilibrium heating nonlinearity, can manifest themselves very differently in terms of their dissipative-reactive ratios and response times. These different characteristics provide a way of experimentally distinguishing the action of different nonlinear mechanisms.

Experimentally, the dissipative-reactive ratio can be deduced from measurements of attenuation and phase shift as functions of input power level; whereas the response time of the underlying nonlinearity can be extracted from measurements of gain as a function of signal frequency. We have fabricated TWPAs based on Al and Ti, and characterised their nonlinearities using our analysis. Our results suggest that for both Al and Ti, the dominant underlying nonlinear mechanism differ significantly from the equilibrium supercurrent nonlinearity. Instead, the measured dissipative-reactive ratios and response times are quantitatively similar to predictions for the non-equilibrium heating nonlinearity.

Currently in the literature, all TWPAs that have successfully reported broad-band amplification are based on alloy or amorphous superconductors such as NbTiN, NbN, TiN, and WSi \citenumns{Eom_2012, Adamyan_2016, Goldstein_2020}. Important research should be performed to investigate the significance of non-equilibrium nonlinear mechanisms in these materials. Interestingly, these materials have also been shown to have quasiparticle lifetimes that are several orders of magnitude smaller compared to that of elemental superconductors \citenumns{Guruswamy_2015, Barends_Lifetime_2008, Janssen_2012, Barends_Lifetime_2008, mazin2020_superconducting}. In the context of this study, two explanations can be offered regarding their success at achieving broad-band amplification: 1) For some materials, the nonlinear behaviour is still dominated by the non-equilibrium heating mechanism. However, the short quasiparticle lifetimes allow the bandwidths of the amplification processes to be in the $\mathrm{GHz}$ range. 2) The onset power of the non-equilibrium heating mechanism increases as the lifetime decreases \citenumns{Rothwarf_1967,Thomas_2020}. When the equilibrium supercurrent nonlinearity has the smaller onset power, the amplifier can be operated at power levels that achieve amplification mainly through the equilibrium supercurrent nonlinearity. Under this operating condition, the fast response time of the equilibrium supercurrent nonlinearity allows the device to achieve broad-band amplification. Theoretical research should be done to quantitatively compare the onset powers of different nonlinear mechanisms using commonly measured material properties, such as critical current density and quasiparticle recombination time. It is valuable to explore design schemes that suppress non-equilibrium nonlinear mechanisms in order to reduce their limitations on the performance of TWPAs.

Beyond the immediate context of TWPAs, our method of characterising nonlinearities can prove useful in the analysis of other superconducting nonlinear devices. Examples of these devices include parametric up-converters \citenumns{Kher_2017}, kinetic inductance Fourier transform spectrometers \citenumns{Faramarzi_2020}, and resonator parametric amplifiers \citenumns{Tholen_2009}. Similar to a TWPA, these nonlinear devices also exploit the underlying superconducting nonlinearities. Dissipative effects and rate-limiting response times are thus likewise important to their operation and performance. A detailed analysis of these effects in the context of resonator parametric amplifiers can be found in \citenumns{Chris_2022_resonator_amplifier}.


\begin{acknowledgements}
The authors are grateful for funding from the UK Research and Innovation (UKRI) and the Science and Technology Facilities Council (STFC) through the Quantum Technologies for Fundamental Physics (QTFP) programme (Project Reference ST/T006307/1), and for funding from the University of Cambridge and the China Scholarship Council (CSC) through the CSC Cambridge Scholarship.
\end{acknowledgements}

\bibliographystyle{h-physrev}
\bibliography{library}
\end{document}